\documentclass[twocolumn,twoside,slac_two]{revtex4}
\usepackage{graphicx}
\usepackage{fancyhdr}
\begin{document}

\title{Theoretical Concepts for Ultra-Relativistic Heavy Ion Collisions}

%

\author{L. McLerran}
\affiliation{RIKEN Brookhaven Center and Brookhaven National Lab., 
Physics Dept.,
Upton, NY 11973-5000 USA}

\begin{abstract}
Various forms of matter may be produced in ultra-relativistic heavy ion collisions.  These are
the Quark Gluon Plasma, the Color Glass Condensate , the Glasma and Quarkyonic Matter.
A novel effect that may be associated with topological charge fluctuations is the Chiral Magnetic 
Effect.  I explain these concepts and explain how they may be seen in ultra-relativistic heavy ion collisions.

\end{abstract}

\maketitle

\thispagestyle{fancy}


\section{Introduction}

The central purpose of ultra-relativistic nuclear collisions is to produce and study forms of high energy 
density strongly interacting matter and to study their properties.  A variety of such forms of matter have been proposed. (References to early developments in this field may be found in Ref. \cite{McLerran:2008qi}) The Quark Gluon Plasma (QGP) is a deconfined phase  of quarks and gluons \cite{Gyulassy:2004zy}. The Color Glass Condensate  (CGC) is high energy density and highly coherent matter made from gluons \cite{Iancu:2003xm}. The Glasma
is matter produced in the first instants of a heavy ion collision from the Color Glass Condensate \cite{Kovner:1995ja}-\cite{Lappi:2006fp}. It involves many strongly interacting and highly coherent color electric and color magnetic longitudinal flux
lines.  Quarkyonic matter is high baryon density matter where the energy density is very high compared to the typical QCD scale, but that still remains confined\cite{McLerran:2007qj}-\cite{Hidaka:2008yy}. I will describe each of these forms of matter and how they might appear in ultra-relativistic nuclear collisions.  

\begin{figure}[h]
\centering
\includegraphics[width=85mm]{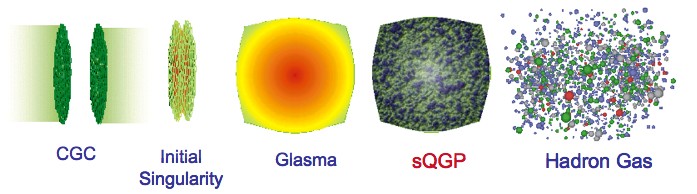}
\caption{Ultra-relativsitc nuclear collisions and various forms of high energy density matter.} 
\label{bass}
\end{figure}

A simple picture of ultra-relativistic nuclear collisions is shown in Fig. \ref{bass} .  Here two Lorentz contracted nuclei, which are thought of as sheets of Color Glass collide.  In the instant these two sheets pass through one another, the fields become singular, and produce the Glasma.  The Glasma fields decay and produce a QGP.  This QGP dilutes through expansion and eventually form a gas composed of ordinary strongly interacting mesons and nucleons.

\begin{figure}[h]
\centering
\includegraphics[width=85mm]{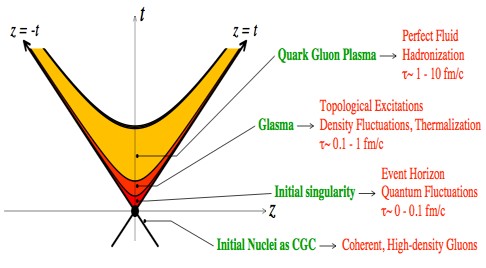}
\caption{A lightcone diagram for ultra-relativistic hadronic collisions.} 
\label{lightcone}
\end{figure}

Sometimes it is useful to conceptualize such collisions in terms of a light cone diagram as shown in Fig. \ref{lightcone}.  This figure shows the various times in the collision.  In particular, in the Glasma
phase, color electric and color magnetic fields produced can carry topological charge.\cite{Kharzeev:1998kz}-\cite{Fukushima:2008xe} In QCD, topological charge fluctuations are associated with the generation of masses for the nucleons.  Analogous processes in electroweak theory may be responsible for the baryon asymmetry of the universe.

There are many theoretical questions that we would like to answer through the collisions of such matter.  Among them are:
\begin{itemize}
\item{What is the high energy limit of QCD?}
\item{What are the possible forms of high energy density matter?}
\item{How do quarks and gluons originate is strongly interacting particles?}
\end{itemize} 

\section{The Color Glass Condensate}

\begin{figure}[h]
\centering
\includegraphics[width=85mm]{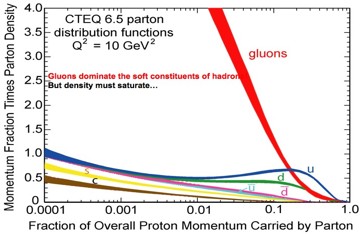}
\caption{The gluon and quark composition of a proton.} 
\label{gluons}
\end{figure}

A remarkable property of strongly interacting particles is that the small x part of their wavefunction is dominated by gluons.  Since $x$ is the ratio of the energy of a constituent to that of the hadron in the reference frame where the hadron is fast moving, the smallest values of $x$ are probed at the highest energies.   It is this gluon rich part of the hadron wavefunction that controls the high energy limit of QCD.
The distribution of gluons as a function of $x$ at $Q^2 = 10 ~GeV^2$ is shown in Fig. \ref{gluons}
for a proton.

Because the size of a proton grows at most like $ln^2(E/E_0)$ at high energies, and because the gluonic contribution to a hadron is measured to grow much faster, like a power $1/x^{\delta}$ where $\delta = 0.2 - 0.3$, the gluons must form a high density state. The coupling evaluated at such a high 
density scale is therefore weak.  Weak coupling does not however mean weak interactions.  Coherence can amplify the effects of an intrinsically weak interaction, as we know is the case in gravity.  

\begin{figure}[h]
\centering
\includegraphics[width=85mm]{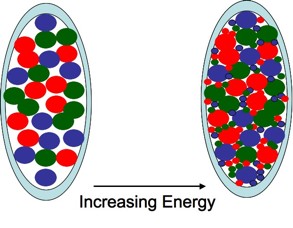}
\caption{Saturation of gluons in the hadron wavefunction.} 
\label{saturation}
\end{figure}

Surely if we try to insert more and more gluons of fixed size into a hadron wavefunction,  at some point it becomes very difficult to add more \cite{Gribov:1984tu}-\cite{Mueller:1985wy}. This is when the interactions of order $\alpha_S$ become strong because there are typically $1/\alpha_S$ particles at that size scale.  This requires that additional particles added to the system are at yet a smaller size scale.  This is shown in Fig. \ref{saturation}  We see that saturation requires that the number of gluons of fixed size will stop rapidly growing above some energy, but the total number of saturated gluons grows forever, since one saturates gluons of smaller an smaller size.   

This high energy density saturated gluonic matter is called the Color Glass Condensate\cite{McLerran:1993ni}-\cite{Ferreiro:2001qy}. The word color arises from the color of the gluons.  The word glass is because the coherent fields at small x are ultimately generated by gluons at larger values of x through evolution equations.  These fast gluons have their natural time scale dilated, and this dilated time scale is then passed on to the fields at small
x. Systems of particles that have long time scales for evolution compared to natural time scales are glasses.  The word condensate is because the phase space density of gluons is very large and they gluons are very coherent.  An effective energy functional in terms of the phase space density,
\begin{equation}
 \rho = {{dN} \over {dyd^2p_Td^2x_T}}
\end{equation}
is
\begin{equation}
  E = -\kappa \rho + \kappa^\prime \alpha_S \rho^2
\end{equation}
The first term in this equation drives the condensation and the second is due to repulsive interactions that stabilize it.  The extremum is when $\rho \sim 1/\alpha_S$.  The dependence on coupling is typical of condensation phenomenon such as Bose condensation, superconductivity, and even the Higgs condensate.

It may seem a little strange to think of a hadron at high energy as a sheet of Color Glass Condensate and at low energies as valence quarks.  It naively violates Lorentz invariance.  The resolution of this apparent paradox is that these correspond to two separate pieces of the hadron wavefunction.  The matrix elements important for high energy scattering are sensitive to the high gluon density part of the hadron wavefunction, where low energy processes are sensitive to the valence quark piece.

The typical density of gluons that arise from the phase space density above is
\begin{equation}
  {{dN} \over {dy}} = \int d^2x_T \int d^2q_T \rho \sim {1 \over \alpha_S} \pi R^2 Q_{sat}^2
 \end{equation}
where $Q_{sat}$ is the maximal momenta for the gluons that are fully saturated, that is where $\rho \sim 1/\alpha_S$.

\section{The Initial Conditions and the Glasma}

\begin{figure}[h]
\centering
\includegraphics[width=85mm]{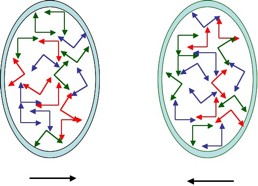}
\caption{The collision of two sheets of colored glass.} 
\label{collision}
\end{figure}

The color electric and color magnetic fields in a sheet of Colored Glass are the Lorentz boosted 
Coulombic fields,  Their color and intensity is random, and they are plane polarized perpendicular to the direction of motion of the hadron.  They are analogous to the Lienard-Weichart potentials of electrodynamics.  The collisions of two nuclei may be thought of as the collision of two sheets of colored glass as shown in Fig. \ref{collision}.  Note that color electric and color magnetic fields are treated on an equal footing, so that the CGC is self-dual under $E \leftrightarrow B$

\begin{figure}[h]
\centering
\includegraphics[width=85mm]{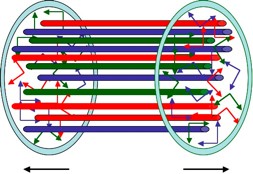}
\caption{Glasma flux tubes.} 
\label{fluxtubes}
\end{figure}

If one takes the initial conditions for the CGC fields, and then computes the field infinitesimally in time after the sheets of color glass pas through one another, one finds longitudinal color electric and color magnetic fields between the two sheets.  It is as if the two sheets of colored glass became dusted with equal and opposite densities of color electric and color magnetic charges.  We can think of the lines of color electric and color magnetic flux that join these sheets as flux tubes.  The typical transverse size of these random flux tubes is $R \sim 1/Q_{sat}$.  Such flux tubes are shown in Fig. \ref{fluxtubes}
This collection of flux tubes and its subsequent evolution is called the Glasma, since it is a transient state of matter that connects the Color Glass Condensate to the Quark Gluon Plasma.  It has coherent fields, like the CGC, but it has properties similar to a plasma with strong background fields, and these fields evlove in time with the natural time scale $1/Q_{sat}$.

Because there are both  longitudinal color electric and  longitudinal color magnetic fields, there is a non-zero density of topological charge, $\vec{E} \cdot \vec{B}$.  A non-zero average of such a quantity would signal
the breaking of P and CP.  Here the average over space is zero, although there are strong local fluctuations.  A topological charge generates non-zero helicity in quarks.  To see this, note that if we accelerate a charge along the direction of $\vec{E}$, it will spiral with a definite handedness around
$\vec{B}$.  If we reverse the charge, the direction reverses and the spiral reverses so the helicity remains the same.

\section{The Quark Gluon Plasma}

\begin{figure}[h]
\centering
\includegraphics[width=85mm]{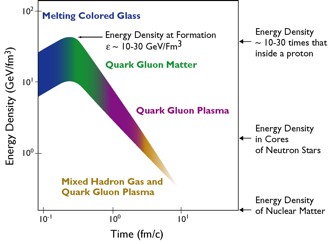}
\caption{Time scales for heavy ion collisions.} 
\label{times}
\end{figure}

Through some as yet not completely understood mechanism, the Glasma eventually thermalizes into a Quark Gluon Plasma.\cite{Baier:2000sb}
The QGP is matter that is deconfined and chirally symmetric \cite{Karsch:2001vs}.
There is evidence that this matter is well thermalized, since the distributions of produced particles are well described by hydrodynamic equations with surprisingly small viscosity\cite{Gyulassy:2004zy},\cite{Teaney:2000cw}, \cite{Arsene:2004fa}-\cite{Adcox:2004mh}. This is only understood if the constituents of the QGP are strongly interacting.  A plot that shows the typical times scales and energy densities believed to be present in RHIC collisions of gold nuclei is shown in Fig. \ref{times}.  Note that the maximal energy density is many times that of the typical scale of QCD $\epsilon \sim 1~ GeV/Fm^3$.

\section{Phenomenological Evidence for the Color Glass Condensate}

The theory of the Color Glass Condensate provides a framework for understanding the high energy limit of QCD.  This framework can be proven from first principles in QCD if the saturation is sufficiently large
so that weak coupling methods can be used.  A substantial question is: When is this true?  

There are a number of pieces of evidence that come from the Hera experiments on deep inelastic scattering, and from fixed target experiments on nuclear targets..\cite{Kowalski:2003hm}-\cite{Kowalski:2008sa} Describing these results involves
knowing the deep inelastic structure function $F_2$,  structure functions for diffractive deep inelastic scattering, and the longitudinal structure function $F_L$.  A simple comprehensive description of these data are possible within the CGC framework \cite{Albacete:2009fh}.

\begin{figure}[h]
\centering
\includegraphics[width=85mm]{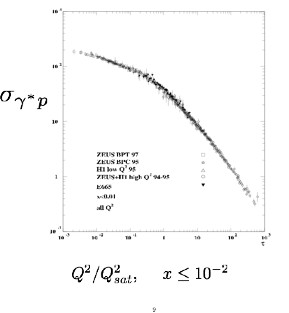}
\caption{Geometric Scaling in Deep Inelastic Scattering} 
\label{geom}
\end{figure}

One measure of the success of the CGC description is its simple prediction of geometric scaling, that is that the cross section for virtual photon scattering from a  protons should scale as $F(Q^2/Q_{sat}^2)(x)$ \cite{Stasto:2000er}-\cite{Munier:2004xu}.
The saturation momentum's dependence on $x$ has been computed and agrees with the fits to the Hera data \cite{Mueller:2002zm}-\cite{Triantafyllopoulos:2002nz}. A geometric scaling curve is shown in Fig. \ref{geom}.

In spite of the remarkably good and simple description that the theory of the CGC gives for the Hera data, there is no consensus within this community that this description is the correct one.  

\begin{figure}[h]
\centering
\includegraphics[width=85mm]{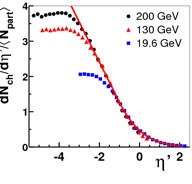}
\caption{Extended scaling in deep inelastic scattering} 
\label{extscaling}
\end{figure}

One of the central results of the theoretical description of the CGC is the existence of a renormalization group description as a function of $x$ or rapidity.\cite{Gelis:2006tb}
 An implication is that there should be an approximate limiting fragmentation or extended scaling.  Such extended scaling has been observed by the Phobos collaboration, shown in Fig. \ref{extscaling}.\cite{Back:2002wb}.
 
One of the earliest successes of the CGC description of heavy ion collisions was its predictions for the total multiplicity of produced particles and its dependence upon the centrality of the collision \cite{Krasnitz:1999wc},\cite{Eskola:2000xq}-\cite{Kharzeev:2000ph}.  Because the saturation momentum provides an infrared cutoff in the computation of particle production,  the total multiplicity is
reliably computed if the saturation momentum is sufficiently large.  It predicts a slow growth with energy and with centrality of the collision for the number of particle produced per nucleon participant in the collision.

The CGC gives a  heuristic derivation of the slow logarithmic growth of total hadronic cross sections
\cite{Kovner:2002yt}-\cite{Ferreiro:2002kv}.
It provides a theory of non-leading twist shadowing, where the saturation momentum appears as the scale factor for the non-leading twist contributions \cite{Baier:2003hr}-\cite{Guzey:2004zp}.   This theory was successfully applied to and predicted the suppression of particles produced at small nuclear $x$ in deuteron gold scattering \cite{Arsene:2003yk}-\cite{Arsene:2004ux} .   
It will be severely tested when data is available from the recent $dAu$ run at RHIC for its predictions concerning $J/\Psi$ productions and forward backward angular correlations for particles in the small
x region of the gold nucleus  .  The prediction for  dAu $J/\Psi$ data is that when the saturation momentum exceeds the charm quark mass, then charm is like a light mass hadron.  It should have approximate  limiting fragmentation, and production cross sections should be reduced relative to incoherent scattering.\cite{Kharzeev:2008nw}  For two particle correlations at small nuclear $x$ , if one triggers on a high transverse momentum particle \cite{Kharzeev:2004bw}-\cite{Qiu:2004id}, then there should be broadening and loss of intensity of the backward going peak relative to pp collisions due to the interactions during the production of the backward going particle.  Leading twist shadowing, that is perturbative QCD, predicts no such suppression.

\section{Predictions of the Glasma}
\begin{figure*}[t]
\centering
\includegraphics[width=135mm]{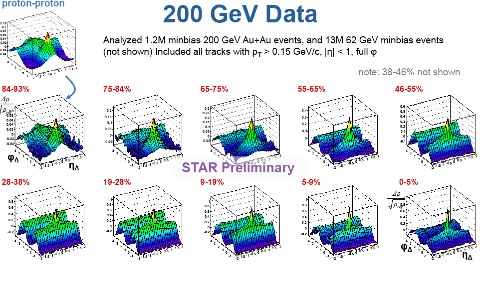}
\caption{The ridge as seen in the STAR experiment} 
\label{starridge}
\end{figure*}
An important feature of the Glasma is long range correlations in rapidity induced by the Glasma flux tubes.\cite{Shuryak:2007fu}-\cite{Gelis:2009tg}. Correlation over a longitudinal momentum scale $p$ must be generated at a time $t \sim 1/p$,
a statement that can be proved using general causality arguments.  In two particle correlation experiments at RHIC, a long range rapidity structure has been observed \cite{Abelev:2009qa}.
The plot is in pseudo-rapidity and azimuthal angle.  It is seen both in two particle correlations where one particle is triggered at high momentum, and inclusively, where there is no minimum momentum requirement.  In Fig. \ref{starridge}, preliminary results on the inclusive ridge from STAR is shown.

\begin{figure}[t]
\centering
\includegraphics[width=85mm]{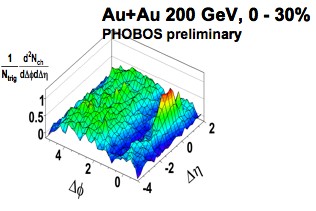}
\caption{The ridge as seen in the Phobos experiment.} \label{phobosridge}
\end{figure}

The data show that for peripheral collisions the two particle correlation is similar to that in pp, where there is no long range component in rapidity.  As the centrality increases a component extended over several units of rapidity appears.  This component has a particle composition similar to single particle inclusive measurements.  The STAR experiment has a limited rapidity coverage.  Over a much longer range in rapidity, the Phobos experiment shows a ridge, as shown in Fig. \ref{phobosridge} \cite{Alver:2009id}.

The ridge can arise from an underlying flux tube structure.  The long range correlation in rapidity is built into the longitudinal color electric and color magnetic flux tubes.  The azimuthal angular correlation can arise form the flux tube flowing with the underlying media in which it was produced.  The flow Lorentz boosts the distribution and narrows an azimuthally symmetric angular distribution in the direction of overall motion of the flux tube.  

The STAR experiment has measured the strength of long range correlations in rapidity. \cite{:2009dq} These increase with the centrality of the collision and for peripheral collisions agree with pp collisions.  For central gold gold collisions at the highest energy, the correlation is as strong as it can be based and is three times stronger than the upper limit predicted by impact parameter fluctuations.  The dependence upon centrality can be understood as trade off of flux tube emission from the Glasma that is leading order in $\alpha_S$ and short range scattering.  In more central collisions or higher energy, $\alpha_S$ is smaller, leading to a relative strengthening of flux tube emission and hence a stronger correlation \cite{Armesto:2006bv}-\cite{Lappi:2009vb}.

The distribution of quanta produced from a single flux tube is a negative binomial distribution.  A negative binomial distribution is parameterized by the number of particles per emitter and the number of emitters.  A sum of negative binomial distributions is a negative binomial distribution.  The Phenix experiments has measured the fluctuation spectrum of produced particle in heavy ion collisions and it is negative binomial.\cite{Adare:2008ns}
 The number of emitters scale as the number of participants, as expected for the Glasma. \cite{Gelis:2009wh}

\section{Topological Charge and the Chiral Magnetic Effect}

The fields in the Glasma carry a topological charge density since there are parallel color electric and color magnetic fields \cite{Kharzeev:1998kz}-\cite{Fukushima:2008xe}.
 Such a distribution has total zero topological charge, but even total topological charge may be generated by evolution of the mater distribution in the Glasma from fixed initial conditions.   Conifigurations with topological charge may generate net helicity for quarks.  In an overall net (electromagnetic) magnetic field,  the magnetic moments of the quarks will align. If there is net helicity, then there is net  motion of the quarks move along the direction of their spins, and an electromagnetic current is set up.  This is the Chiral Magnetic Effect. 

\begin{figure}[t]
\centering
\includegraphics[width=85mm]{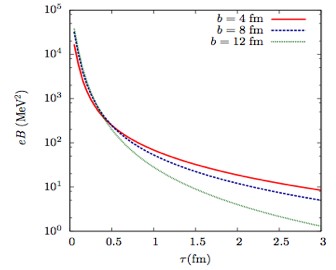}
\caption{The typical magnetic field strength produced in the overlap region in the collision of two 
Gold nuclei.} 
\label{magfield}
\end{figure}

In off impact parameter zero heavy ion collisions a net magnetic field is set up in the region where the nuclei overlap.  Such a field decays away quickly in time and can be computed from knowledge of the distributions of ahcarged particle produced in such collisions.  The result of a computations of this field is shown in Fig. \ref{magfield}.  It is quite intense at early times and can have a characteristi scale of order $100~MeV$.  If there is a topological charge fluctuation that generates a net helicity, this will therefore appear as a fluctuation in the net charge of particles perpendicular to the reaction plane.

\begin{figure}[t]
\centering
\includegraphics[width=85mm]{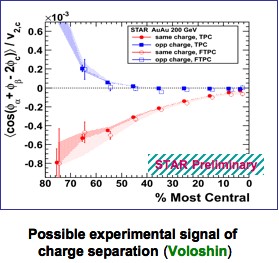}
\caption{Fluctuations in charge orthogonal to the reaction plane.} 
\label{startop}
\end{figure}

Of course any effect  in QCD must be P and CP odd, so effects generated by the chiral magnetic effect must average to zero.  Nevertheless, their fluctuation spectra nay be predicted by QCD, but they must be carefully compared against other possible effects.  One measurable is the fluctuations in particle perpendicular to the reaction plane.  This has been measured in STAR and is shown in Fig. \ref{startop}  \cite{Voloshin:2009hr}.

The results shown agree semi-quantitatively with predictions of the chiral magnetic effect
\vspace{0.1in}

\section{The Decay of the Glasma and Thermalization}

The Glasma configuration will decay away into gluonic quanta.  This decay is chaotic and turbulent 
\cite{Mrowczynski:2004kv}-\cite{Rebhan:2004ur} Two field configurations initially very similar deviate from one another exponentially in time.  This is due to modes dependent on rapidity.  The lowest order Glasma solution is taken to be invariant in rapidity, but the fluctuations will generate deviations from this.  This has led to a number of speculations about turbulent thermalization  Such thermalization might lead to anomalously low viscosities as suggested by the RHIC data.  It might also lead to significant jet energy loss at early times as jets traverse the Glasma \cite{Asakawa:2006tc}-\cite{Shuryak:2002ai}.

The problem with all of these speculations is that the Glasma lives a very short time $t \sim 1/\alpha_S Q_{sat}$ before the effects of any coherent field should be drained away.  On the other hand, thermalization involves scattering, and we would expect time scales of roder $t \sim 1/\alpha_S^2 Q_{sat}$ for this to occur.  If the matter is thermalized, it has little coherent component, and the description of distributions of particles and flow using hydrodynamic simulation suggests that such matter has very small viscosity, and hence is very strongly interacting.

Another possibility is that the thermalization is a consequence of the dynamics becoming very strongly coupled as the matter expands.  This strongly coupled quark gluon plasma may have qualitative features in common with that of the AdSCFT model of QCD interactions.  This theory is $N = 4$ supersymmetric Yang-Mills theory and as such has a peripheral relation to QCD.  The theory predicts
a lower bound on the viscosity to entropy ratio that is a factor of 2-3 times smaller than that which seems to describe experimental data within hydrodynamic simulation. \cite{Kovtun:2004de}
The lower limit found in AfSCFT computations is close to that found by uncertainty principle bounds, that is, the mean free path must be larger than the Compton wavelength of a particle.

AdSCFT computations have proved useful in a more theoretical context:  The naive extrapolation of perfect fluid hydrodynamics to include viscous terms violates either causality, or the requirement that entropy is always increasing as a function of time.  The AdSCFT computations suggested specific forms for higher order modifications of the hydrodynamic equations that do not violate these principles, and provide estimates for the magnitude of coefficients that characterize these corrections \cite{Romatschke:2007mq}-\cite{Betz:2009zz}.

One of the outstanding problems in the description of the QGP is a quantitative evaluation of jet energy loss.\cite{Gyulassy:2000fs}
Jets are suppressed by a factor of 4-5 in central Au-Au collisions at RHIC energies out to an energy of at lest $15~GeV$.  Heavy quarks such as charm, are suppressed to the same degree as
light quarks, and appear to flow with the media produced like light quarks.  These phenomena are difficult but possible to include in weak coupling QCD computations.

Perhaps a more natural explanation arises in the AdSCFT strong coupling world.  The problem is that in this strongly couple world, jets are produced neither in electron-positron annihilation nor in deep inelastic scattering.  If one tries to produce a jet, it immediately produces particle and cascades down to low momentum, producing isotropic distributions of low momentum quanta.  This has the consequence that the distribution of parton in a hadron scale as $1/x^2$, and all the energy in a hadron is in the low $x$ quanta.  In fact these quanta saturate, and one makes a strongly couple Color Glass Condensate.
These features may resemble low energy strong interactions more than high energy, and for the reasons stated is difficult to apply to jet quenching.\cite{Hatta:2007he}-\cite{Hatta:2008tx}

\begin{figure*}[t]
\centering
\includegraphics[width=135mm]{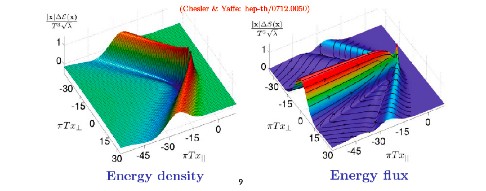}
\caption{A Mach cone generated by a fast moving particle in an AdSCFT computation} 
\label{mach}
\end{figure*}

Nevertheless, the experimental data on jet energy loss and heavy particle flow is stunning.  This is particularly so for the distribution of particles produced by the backward going partner of a tagged jet..
The distribution of such particles  is double peaked around 180 degrees opposite in azimuthal angle  of the tagged jet.   This has led to speculations that the effect may arise from a Mach cone, that is the backward going jet propagating supersonically in the QGP media.  Such Mach cones are clearly seen in the AdSCFT simulations, Fig. \ref{mach}  \cite{Gubser:2006qh}-\cite{Torrieri:2009mv}.
 
\vspace{0.1in}

\section{Quarkyonic Matter}

Matter at high baryon density was the subject of studies at the CERN SPS and the AGS.  There will soon be a study using low energy beams at RHIC.  There have been a number of speculations concerning
possible Color Superconductivity of this matter, that may occur at very high baryon densities and quite low temperature.\cite{Alford:1997zt}-\cite{Rapp:1997zu}  Recent arguments have suggested that the phase structure of QCD matter is even richer than had previously been envisioned.\cite{McLerran:2007qj}-\cite{McLerran:2008ua}

QCD matter was originally thought to be divided into a confined word of baryons and mesons and a deconfined world of quarks and gluons.  The new feature on this phase diagram is Quarkyonic Matter.
This matter is confined, but the Fermi sea of quarks has interactions that can be treated perturbatively.
Baryon excitation near the Fermis surface. and all mesonic and glueball excitations are confined.  The word quarkyonic is a hybrid of quark and baryonic for this reason.

The reason why quarkyonic matter is confined is that  at low temperature, the only particles that can short out the quark potential are quark loops.  In the large $N_c$ limit of QCD, these loops are suppressed by a factor of $1/N_c$.  So at least at large $N_c$, there is a world that has densities very high compared to the QCD scale $\Lambda_{QCD}^3$ that is confined. 

In so far as the large $N_c$ limit is valid, the transition to the deconfined world should be indepedant of density.  This is confirmed in lattice computations where the deconfinement temperature depends only weakly on baryon density.

The phase transitions in QCD are largely characterized by changes in the number of degrees of freedom.  The confined world has light mass pion degrees of freedom.  The decconfined world has
$2(N_c^2-1)$ gluon and $4N_cN_F$ quark-antiquark degrees of freedom.  The quarkyonic world has the degrees of freedom of $2N_cN_f$ quarks  plus light mass meson degrees of freedom.

The chiral properties of the quarkyonic world are not yet understood although there have been some interesting and strongly motivated conjectures\cite{Glozman:2007tv}-\cite{Glozman:2008jw}.

A prominent feature of the QCD phase diagram is a triple point where the quarkyonic,
confined and deconfined world coexist.  There may be a nearby critical point if the transitions between these different phases for some range of temperature and density are truly phase transitions, as opposed to cross overs.  If there is no true phase transition, then one expects rapid cross over between the various phases. 

\begin{figure}[t]
\centering
\includegraphics[width=85mm]{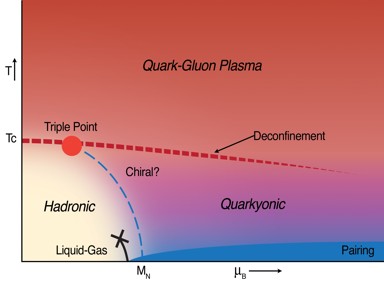}
\caption{The phases of QCD at finite temperature and density.} 
\label{quarkyonic}
\end{figure}

\section{Conclusions}

The strongly interacting Quark Gluon Plasma is well described by perfect fluid hydrodynamic computations at RHIC.  The origin of this behaviour, and jet quenching within the Quark Gluon Plasma has proven difficult to understand.

The Color Glass Condensate are suggested by the data from RHIC and Hera but their is not yet consensus that this is the origin of the various phenomena elaborated in the paper.  I believe a compelling case can be made after the preliminary data on the ridge and long range correlations is published, and data on forward-backward correlations at forward rapidiities is represented and published.

Mater at high baryon density is less understood.  This is a difficult issue to treat theoretically since numerical Monte Carlo methods that conclusively demonstrated the properties of matter at high temperature do not work well at high baryon density.  Experimental data on the high baryon density region is difficult since this involves experiments at lower energy, and it is not clear how far one can probe the high density region.

Please note that this  paper contains a review of recent theoretical developments in the theory of high energy density strongly interacting matter.  It is not complete in its treatment, and  in particular the discussion of more established ideas in the field.  It reflects what I consider to be of interest for many theorists in the field.  As such, and since this is a conference talk, not an extensive review, the references are not as comprehensive as I might have preferred.

\vspace{0.1in}

\begin{acknowledgments}

The research of  L. McLerran is supported under DOE Contract No. DE-AC02-98CH10886.
\end{acknowledgments}

\end{document}